# Pressure-induced coevolution of transport properties and lattice stability in CaK(Fe$_{1-x}$Ni$_x$)$_4$As$_4$ (x= 0.04 and 0) superconductors with and without spin-vortex crystal state

Pengyu Wang[1,2]*, Chang Liu[1,2]*, Run Yang[1]*, Shu Cai[1,3], Tao Xie[1,2], Jing Guo[1,5], Jinyu Zhao[1,2], Jinyu Han[1,2], Sijin Long[1,2], Yazhou Zhou[1], Yanchun Li[3], Xiaodong Li[3], Huiqian Luo[1,2,5], Shiliang Li[1,2,5], Qi Wu[1], Xianggang Qiu[1,2,5], Tao Xiang[1,2], and Liling Sun[1,2,3,5]†

[1]Institute of Physics, Chinese Academy of Sciences, Beijing 100190, China
[2]University of Chinese Academy of Sciences, Beijing 100190, China
[3]Center for High Pressure Science & Technology Advanced Research, 100094 Beijing, China
[4]Institute of High Energy Physics, Chinese Academy of Science, Beijing 100049, China
[5]Songshan Lake Materials Laboratory, Dongguan, Guangdong 523808, China

Here we report the first investigation on correlation between the transport properties and the corresponding stability of the lattice structure for CaK(Fe$_{1-x}$Ni$_x$)$_4$As$_4$ (x=0.04, 0), a new type of putative topological superconductors, with and without a spin-vortex crystal (SVC) state in a wide pressure range involving superconducting to non-superconducting transition and the half- to full-collapse of tetragonal (h-cT and f-cT) phases, by the complementary measurements of high-pressure resistance, Hall coefficient and synchrotron X-ray diffraction. We identify the three critical pressures: $P_{ch}^{on}$ that is the turn-on critical pressure of the h-cT phase transition and it coincides with the critical pressure for the sign change of Hall coefficient from positive to negative, a manifestation of the Fermi surface reconstruction; $P_{ch}^{off}$ that is the turn-off pressures of the h-cT phase transition; and $P_{cf}$ that is the critical pressure of the f-cT phase transition. By comparing the high-pressure results measured from the two kinds of samples, we find a distinct "left-shift" of the $P_{ch}^{on}$ for the doped sample, at the pressure of which its SVC state is fully suppressed, however the $P_{ch}^{off}$ and the $P_{cf}$ remain the same as that of the undoped one. Our results not only provide a consistent understanding on the results reported before, but also demonstrate the importance of the Fe-As bonding in stabilizing the superconductivity of the iron pnictide superconductors through the "pressure window".

The discovery of the high-$T_c$ iron-pnictide and iron-selenide superconductors [1, 2] provides a new platform to specify the essential ingredients existed in cuprate superconductors, and an opportunity to achieve a better understanding on the mechanism of high-$T_c$ superconductivity, a 'holy grail' in the field of the contemporary condensed matter physics and material sciences [3-5]. Now, several types of iron-pnictide superconductors have been found [6-9], all of which have the tetrahedra structure with the Fe-As layers stacking alternatively with other intermediary layers/atoms. Growing evidence from experiments indicates that the Fe-As layers play a key role in developing and stabilizing superconductivity [9]. In 2016, a new member of the iron pnictide superconductors, AeAFe$_4$As$_4$ (Ae=Ca, Sr and A= K, Rb, Cs), has been found [10-26]. Structurally, it can be regarded as the hybrid phases between AeFe$_2$As$_2$ and AFe$_2$As$_2$ with the stoichiometric composition, while is different from most of the other known high-$T_c$ superconductors whose superconductivity are induced by chemical doping. As a result, these stoichiometric superconductors have no chemical substitution-induced inhomogeneity on the lattice, and significantly reduced the complexity of the local structure for understanding the physics behind. Since the Ae$^{2+}$ and A$^{1+}$ atoms are inserted alternately across the Fe$_2$As$_2$ layers, the AeAFe$_4$As$_4$ superconductors host two different sites of As ions in a unit cell (As(1) and As(2), respectively) in the tetragonal unit cell with group space *P*4/*mmm* [10]. Upon cooling, they show a superconducting transition at the temperature ($T_c$) varying from 31 to 36 K. Partial substituting Fe with Ni, Co or Mn reduces the $T_c$ value by more than 10 K due to the existence of a spin-vortex crystal (SVC) state, with Fe spins lying in-plane and stacking along *c* axis antiferromagnetically, that competes with the superconducting (SC) state [22-24]. Remarkably, recent angle-resolved photoemission spectroscopy and scanning tunneling microscopy/spectroscopy experiments find the evidence of the Dirac surface state and Majorana zero mode in the $CaKFe_4As_4$ superconductor [21], implying that this type of superconductors may host some nontrivial quantum states, such as topological superconductivity. Moreover, inelastic neutron scattering experiments find that the spin resonance of the $CaK(Fe_{1-x}Ni_x)_4As_4$ superconductor displays both odd and even modes along the *L* direction [17,18], similar to the resonance observed in bilayer cuprate superconductors [27]. These findings attract additional research interest on them.

Pressure tuning is an effective and clean way to manipulate the crystal and electronic structures without changing the chemistry, often providing significant information for understanding the underlying physics of the exotic state emerging from ambient pressure materials, through investigating the coevolution of electronic states and crystal structure. Some high-pressure studies on the $AeAFe_4As_4$ superconductors have been performed, and interesting results have been achieved from both experimental and theoretical sides [12-16, 26], including the transition of bulk-to-percolating superconducting state [12], half-to-full collapse of the tetragonal phase [14, 15, 26]. These results provide the fundamental knowledge that the superconductivity and the lattice structure of these superconductors are intimately correlated and sensitive to the external pressure. However, the reported investigations of transport properties on Ni-doped and undoped CaKFe$_4$As$_4$ superconductors are performed below the pressure of ~ 6 GPa. What happens under the pressure above 6 GPa remains unclear. In addition, the experimental evidence for the change of electronic state around the critical pressure of the half collapse of the tetragonal (h-cT) phase is still lacking. In this work, we perform the high-pressure studies on the doped and undoped sample up to the pressure where the superconductivity is fully suppressed and the f-cT phase forms.

High-quality single crystals of CaK(Fe$_{1-x}$Ni$_x$)$_4$As$_4$ (x=0.04 and 0) were grown using the self-flux method [17, 18, 25]. The ambient-pressure values of $T_c$s of the doped- and undoped-samples were determined to be 21 K and 35 K, respectively, and the SVC transition for the x=0.04 sample is about $T_N$=44 K [24].

High-pressure resistance measurements were performed in a diamond-anvil cell (DAC), in which diamond anvils with 300 $\mu$m culets (flat area of the diamond anvil) and a nonmagnetic rhenium gasket with 100-$\mu$m-diameter hole were adopted. The standard four-probe electrodes were applied on the cleavage plane of the CaK(Fe$_{1-x}$Ni$_x$)$_4$As$_4$ (x=0.04 and 0) crystals. To obtain a quasi-hydrostatic pressure environment for the samples, NaCl powder was employed as the pressure medium. High-pressure Hall coefficient measurements were carried out by the van der Pauw method. To keep the sample in the same pressure environment as that in the resistance measurements, NaCl powder was also employed as the pressure medium. High-pressure X-ray diffraction (XRD) measurements on CaK(Fe$_{0.096}$Ni$_{0.04}$)$_4$As$_4$ were performed at room temperature on beamline 4W2 at the Beijing Synchrotron Radiation Facility. Diamonds with low birefringence were selected for the measurements. A monochromatic X-ray beam with a wavelength of 0.6199 Å was employed. The pressure for all measurements was determined by the ruby fluorescence method [28].

We first performed temperature-dependent resistance measurements on the single crystal of $CaK(Fe_{0.096}Ni_{0.04})_4As_4$ in a DAC. To investigate the doping effect on the transport properties, a parallel measurement was conducted on the undoped $CaKFe_4As_4$ sample. As shown in Fig.1a and 1c, the superconductivity of both samples with and without the SVC state is sensitive to the pressure applied. For the sample with the SVC state, the ambient-pressure $T_c$ is much lower than that of the sample without the SVC state. It shows that application of pressure renders $T_c$ decreased, no matter whether they host the SVC state or not. Intriguingly, the superconductivity of these two samples is fully suppressed at almost the same critical pressure (~ 11 GPa), as shown in Fig.1b and 1d, suggesting that the initial SVC state has little influence on the critical pressure for destroying the superconducting state. We repeat the measurements with several new samples and obtain the reproducible results (see SI - Ref.[29]). Moreover, we investigate the pressure effect on the onset transition temperature ($T_N$) of the SVC state for our $CaK(Fe_{0.96}Ni_{0.04})_4As_4$ samples and find that $T_N$ exhibits a monotonous decrease with the increment of pressure (see Ref.[29]), similar to what is observed in $CaK(Fe_{1-x}Ni_x)_4As_4$ (x=0.033 and 0.05) and $CaK(Fe_{1-x}Mn_x)_4As_4$ (x=0.024) samples [13, 16].

To identify the similarity and peculiarity in the transport properties, superconductivity and the electronic state in $CaK(Fe_{0.096}Ni_{0.04})_4As_4$ and $CaKFe_4As_4$ samples, we performed high-pressure measurements on Hall resistance ($R_{xy}$) by sweeping the magnetic field ($H$) applied perpendicular to the *ab* plane of the samples, from 0 T to 6 T at 30 K for $CaK(Fe_{0.096}Ni_{0.04})_4As_4$ (Fig.2a and 2b) and at 40 K for $CaKFe_4As_4$ (Fig.2c and 2d). We find that $R_{xy}(H)$ is positive below 2.4 GPa (the average value of the two independent runs - 2.33 GPa+2.46 GPa)/2=2.4 GPa) for the $CaK(Fe_{0.096}Ni_{0.04})_4As_4$ samples, and 3.88 GPa for the $CaKFe_4As_4$ samples (also the average value - (3.90 GPa+3.85 GPa)/2=3.88 GPa). The plots of $R_{xy}(H)$ from our samples indicate that a hole-electron carrier balance ($R_{xy}(H)=0$) occurs at a critical pressure ($P_c$). Although the value of the $P_c$ is different for the samples with and without the SVC state, the behavior of sign change in $R_H (P)$ is the same - below the $P_c$, the sample is dominated by hole carriers, while above the $P_c$, the sample is dominated by electron carriers. The observation of the pressure-induced sign change of the Hall coefficient at the $P_c$ provides important experimental evidence for the dramatic change of electronic structure - the reconstruction of the Fermi surface from a hole dominated to an electron dominated ones [30, 31].

Since the structural stability is one of the key issues for understanding the phenomena found in the pressure range of our experiments, we perform the high pressure X-ray diffraction (XRD) measurements on the $CaK(Fe_{0.096}Ni_{0.04})_4As_4$ sample for the first time. The XRD patterns collected at different pressures are shown in Fig. 3a. It is seen that all peaks measured under pressure up to 39.3 GPa can be well indexed by the tetragonal phase in the *P/4mmm* space group, indicating that no structure phase transition occurs in the pressure range investigated. In Fig.3b, we illustrate the crystal structure of the CaK(Fe$_{0.096}$Ni$_{0.04}$)$_4$As$_4$ sample and define the As ions adjacent to the Ca layers as As(1) and the K layers as As(2).

However, the lattice parameter *a* extracted from our XRD data shows an increase starting at ~ 2.5 GPa and reaching the maximum at ~ 5 GPa, meanwhile the lattice parameter *c* displays a rapid decrease in this pressure range (Fig.4a). These results lead us to propose that the transition of the half-collapse of the tetragonal (h-cT) phase turns on and off at ~ 2.5 GPa and ~5 GPa, respectively (Here we define these critical pressures as $P_{ch}^{on}$ and $P_{ch}^{off}$), and the initial tetragonal (T) phase and the h-cT phase coexist in this pressure range. These results are similar to what have been observed in the pressurized CaKFe$_4$As$_4$ samples, in which the lattice parameter *a* begins to increase at ~3.5 GPa and reaches a maximum at ~ 4.7 GPa [12]. Upon further compression to ~11 GPa, the second collapse occurs - the lattice parameter *a* and *c* also appear noticeable changes (Here we define this critical pressure as $P_{cf}$), implying that the T phase fully collapses (due to lack of more experimental information on the change of the f-cT phase, we are not able to identify the critical pressure that turns off the f-cT phase). The pressure-induced two collapses in the tetragonal CaK(Fe$_{0.096}$Ni$_{0.04}$)$_4$As$_4$ sample are in accordance with the theoretical calculations and experimental results obtained from the measurements on the CaRbFe$_4$As$_4$ and Cs/RbEuFe$_4$As$_4$ samples [14, 15, 26].

We summarize our high-pressure results obtained from the measurements on

CaK(Fe$_{0.096}$Ni$_{0.04}$)$_4$As$_4$ in the pressure-$T_c$ phase diagram (Fig.4b). The four distinct regions defined by the lattice structure can be seen in the diagram: (1) Low-pressure T phase region below $P_{ch}^{on}$, in which the SVC state coexists with the SC state. When pressure is applied, both of $T_N$ and $T_c$ decrease with increasing pressure until the pressure reaches to the $P_{ch}^{on}$. (2) The coexisted T and h-cT phase region that lies in the range of $P_{ch}^{on}$ and $P_{ch}^{off}$. At the $P_{ch}^{on}$, the SVC state is entirely suppressed and the sample starts the transition from a T phase to h-cT phase (Fig. 4a). Just at this pressure, the Hall coefficient ($R_H$) changes its sign from the positive to the negative (Fig. 4c). These results demonstrate that the pressure drives a reconstruction of Fermi surface which is associated to the transition from the T phase to the h-cT phase in a pressure range below ~5 GPa (Fig.4a). (3) The h-cT phase region lied in the range of $P_{ch}^{off}$ and $P_{cf}$, in which the $T_c$ decreases continuously and disappears at the pressure of ~ 11 GPa. (4) The f-cT phase region above the critical pressure of $P_{cf}$, in which the h-cT phase fully converts to the f-cT phase, and the corresponding electronic state of the h-cT phase is taken over by that of the non-superconducting f-cT phase.

To clarify the doping effect on the high-pressure behavior found in the sample with the SVC state, we compare its high-pressure experimental results with that measured from the undoped CaKFe$_4$As$_4$ sample. As shown in Fig.4d-4f, CaKFe$_4$As$_4$ bears the similar high-pressure behavior to that of CaK(Fe$_{0.096}$Ni$_{0.04}$)$_4$As$_4$: $T_c$ monotonically declines with the increment of pressure (Fig.4e), in agreement with the results reported in [12]. At ~3.88 GPa ($P_{ch}^{on}$), the Hall confident ($R_H$) also displays a sign change from the positive to the negative, indicating that the ambient-pressure T phase of the sample begins its half collapse at this pressure [12,14]. A sudden increase of the lattice parameter $a$ and the volume drop are also observed at ~ 3.5 GPa by Kaluarachchi *et al* [12], very close to the $P_{ch}^{on}$ determined by our Hall coefficient measurements (Fig.4f). Since the observed $P_{ch}^{on}$ (~3.88) and $P_{ch}^{off}$ (~ 4.9 GPa) by our Hall coefficient measurements is close to these (~3.5 GPa and ~4.7 GPa) determined by the high pressure and low temperature XRD measurements on the same sample [12], we define the pressures of 3.88 GPa and 4.9 GPa as the $P_{ch}^{on}$ and $P_{ch}^{off}$ of our undoped sample. On further increasing pressure, $T_c$ measured from both CaK(Fe$_{0.096}$Ni$_{0.04}$)$_4$As$_4$ and CaKFe$_4$As$_4$ samples shows a monotonously decrease till the pressure around $P_{cf}$ =11 GPa (Fig.4b a and 4e), at the pressure of which doped- and undoped-samples undergo a transition from the h-cT phase to the f-cT phase [12, 26] and lose their superconductivity. This is the first report on the observation of the pressure-induced SC-to-NSC transition and the identification on that the critical pressure of the SC-NSC transition coincides with the pressure of the lattice transition from the h-cT phase to the f-cT phase in the CaK(Fe$_{1-x}$Ni$_x$)$_4$As$_4$ (x=0.04 and 0) superconductors.

Theoretical calculations on the undoped system find that the formation of the As(1)-As(1) bond across the Ca layers and the formation of the As(2)-As(2) bond across the K layer are responsible for the presence of the h-cT and the f-cT phases [12, 26,32], and propose that the formed As-As bond weakens the Fe-As bonding [32-34], which, in turn, greatly affects the stability of superconductivity [32]. Our findings in the CaK(Fe$_{1-x}$Ni$_x$)$_4$As$_4$ (x=0.04 and 0) superconductors experimentally provides a strong support for the prediction that Fe-As bonding is one of the essential ingredients for the presence of superconductivity in this kind of iron pnictides.

By comparing the high-pressure behavior of the CaK(Fe$_{0.096}$Ni$_{0.04}$)$_4$As$_4$ and CaKFe$_4$As$_4$ samples, we find that the substitution of Ni on Fe site shifts the $P_{ch}^{on}$ to lower pressure, while it has no obvious effects on the $P_{ch}^{off}$ and the $P_{cf}$. Thus, understanding why the existence of the SVC state renders the $P_{ch}^{on}$ to shift to lower pressure should be one of the key issues to reveal the underlying physics of the lattice half collapse. We suggest that the left shift of the $P_{ch}^{off}$ in the Ni-doped sample may be attributed to the interplay between the SVC state and the Fe-As hybridized state (the hybridization between the Fe 3$d$ orbital and As 4$p$ orbital electrons), which makes the hybridized band unstable and benefits the formation of the As(1)-As(1) bonding. These results imply that the $P_{ch}^{on}$ is sensitive to the existence of the competing order introduced by the chemical doping. While the unchanged $P_{ch}^{off}$ and $P_{cf}$ is possibly related to the stability of the lattice that is still governed by the matrix of the initial T phase.

In conclusion, the coevolution of the SVC state, superconductivity, dominated carriers and stability of lattice structure in CaK(Fe$_{1-x}$Ni$_x$)$_4$As$_4$ (x=0.04 and 0) superconductors has been investigated by the complementary measurements of high-pressure resistance, Hall coefficient and synchrotron X-ray diffraction in a wide pressure range involving superconducting to non-superconducting transition, and the corresponding half- to full-collapse of tetragonal phase transition, for the first time. We identified the three critical pressures though the comprehensive analysis of our results: $P_{ch}^{on}$ and $P_{ch}^{off}$ that are the turn-on and -off pressure of the h-cT phase transition, respectively, and $P_{cf}$ that is the critical pressure of the f-cT phase transition. Our results demonstrate that the formation of the As(1)-As(1) bond that is evidence by the transition from the T phase to the h-cT phase changes the sign of $R_H$, a manifestation for the reconstruction of the Fermi surface. While, the formation of the As(2)-As(2) bond characterized by the f-cT phase transition terminates the superconductivity. These results achieved in this study not only provide a consistent understanding on the results reported before, but also demonstrate the importance of the Fe-As bonding in stabilizing the superconductivity of the iron-pnictide superconductors through the "pressure window".

**Acknowledgements**

This work was supported by the National Key Research and Development Program of China (Grant Nos. 2021YFA1401800, 2022YFA1403900, 2018YFA0704200), the National Natural Science Foundation of China (Grant Nos. U2032214, 12104487, 12122414, 12004419, 11822411 and 11961160699), and the Strategic Priority Research Program (B) of the Chinese Academy of Sciences (Grant No. XDB25000000). J. G., S.C. and H. L. are grateful for supports from the Youth Innovation Promotion Association of the Chinese Academy of Sciences (Grant Nos. 2019008, Y202001) and the China Postdoctoral Science Foundation (E0BK111).

These authors with star (*) contributed equally to this work.

Correspondence and requests for materials should be addressed to L.S. (llsun@iphy.ac.cn)

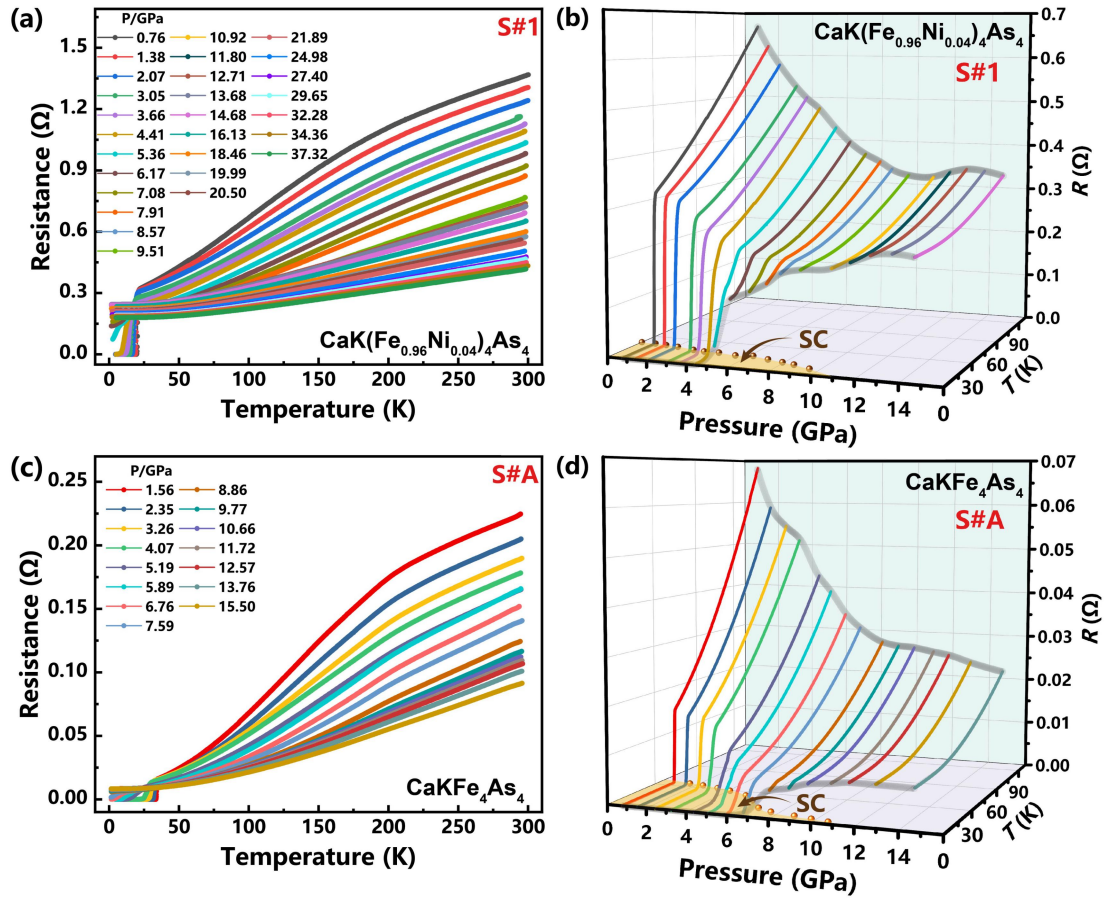

**Figure 1 The results of resistance measurements on the $CaK(Fe_{1-x}Ni_x)_4As_4$ (x=0.04 and 0) at high pressures.** (a) Temperature dependence of the resistance in the pressure range of 0.76 GPa–37.32 GPa for the sample #1 of the $CaK(Fe_{0.96}Ni_{0.04})_4As_4$ single crystal. (b) Enlarged views of the resistance-temperature curves at different pressures for the sample #1 of $CaK(Fe_{0.96}Ni_{0.04})_4As_4$. (c) Resistance as a function of temperature for pressures ranging from 1.56 GPa to 15.5 GPa for the sample #A of $CaKFe_4As_4$ single crystal. (d) Resistance-temperature curves at different pressures for the sample #A of $CaKFe_4As_4$.

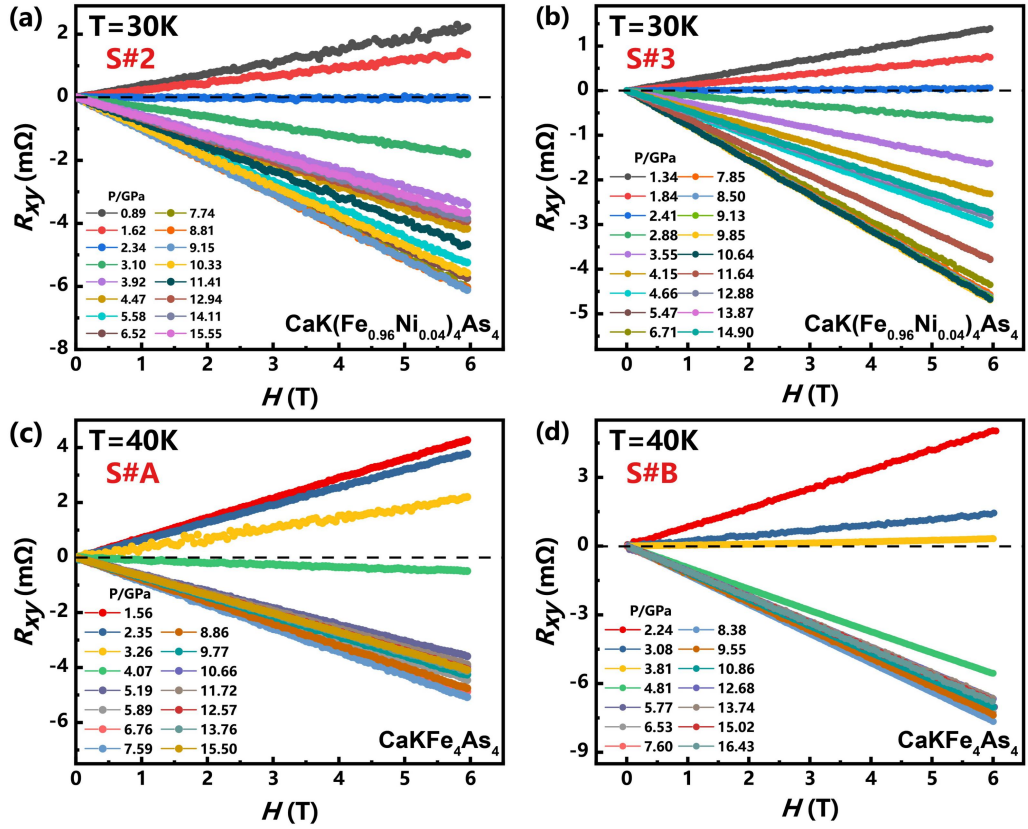

**Figure 2 Hall resistance ($R_{xy}$) as a function of magnetic field ($H$) for the CaK(Fe$_{1-x}$Ni$_x$)$_4$As$_4$ (x=0.04 and 0) single crystals**. Plots of $R_{xy}$ versus $H$ for the CaK(Fe$_{0.96}$Ni$_{0.04}$)$_4$As$_4$ single crystals measured at 30 K in the pressure range of (a) 0.89-15.5 GPa, (b) 1.34-14.9 GPa. $R_{xy}$ versus $H$ for the CaKFe$_4$As$_4$ single crystals measured at 40 K in the pressure range of (c) 1.56 – 15.5 GPa, (d) 2.24-16.43 GPa. The dashed line indicates $R_{xy}(B)=0$ where the average pressures for the sample #2 and #3 of the CaK(Fe$_{0.96}$Ni$_{0.04}$)$_4$As$_4$ and the sample #A and #B of the CaKFe$_4$As$_4$ single crystals are estimated to be ~ 2.4 GPa and 3.88 GPa, respectively.

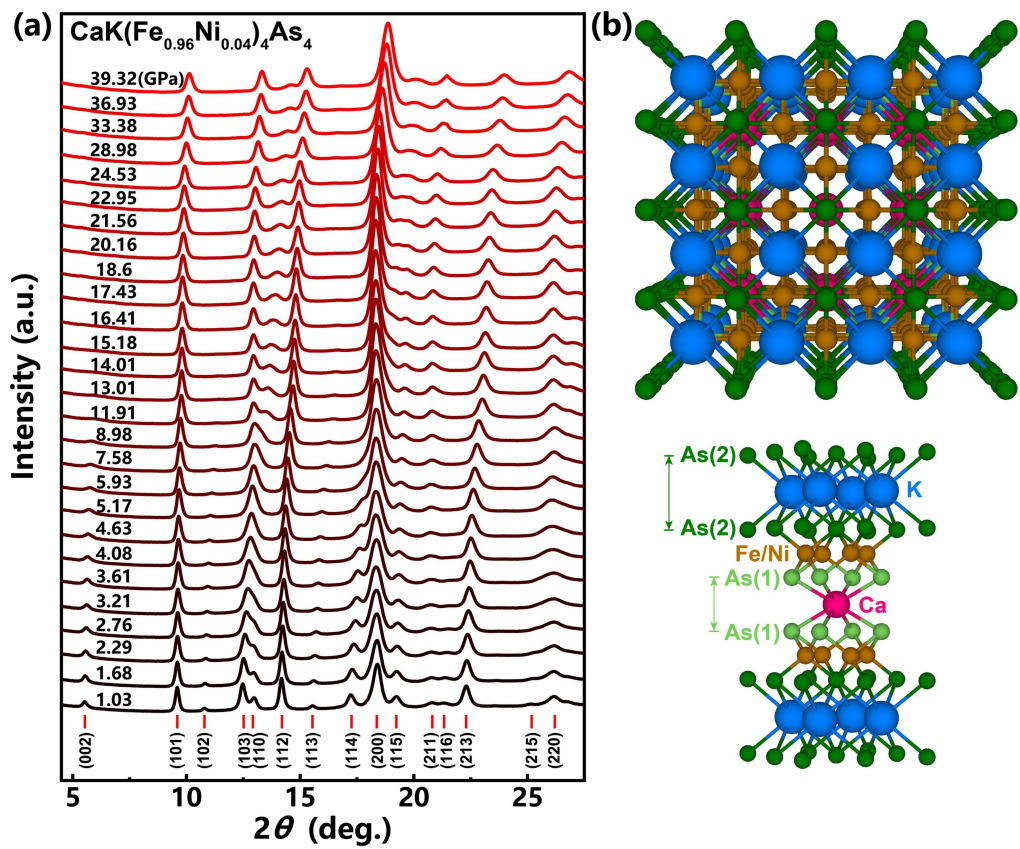

**Figure 3. X-ray diffraction results of the CaK(Fe$_{0.96}$Ni$_{0.04}$)$_4$As$_4$ sample collected at high pressure and crystallographic illustration.** (a) X-ray diffraction patterns measured at different pressures, showing no crystal structure phase transition in the experimental pressure range up to 39.32 GPa. (b) Schematic crystal structure of the CaK(Fe$_{0.96}$Ni$_{0.04}$)$_4$As$_4$ sample.

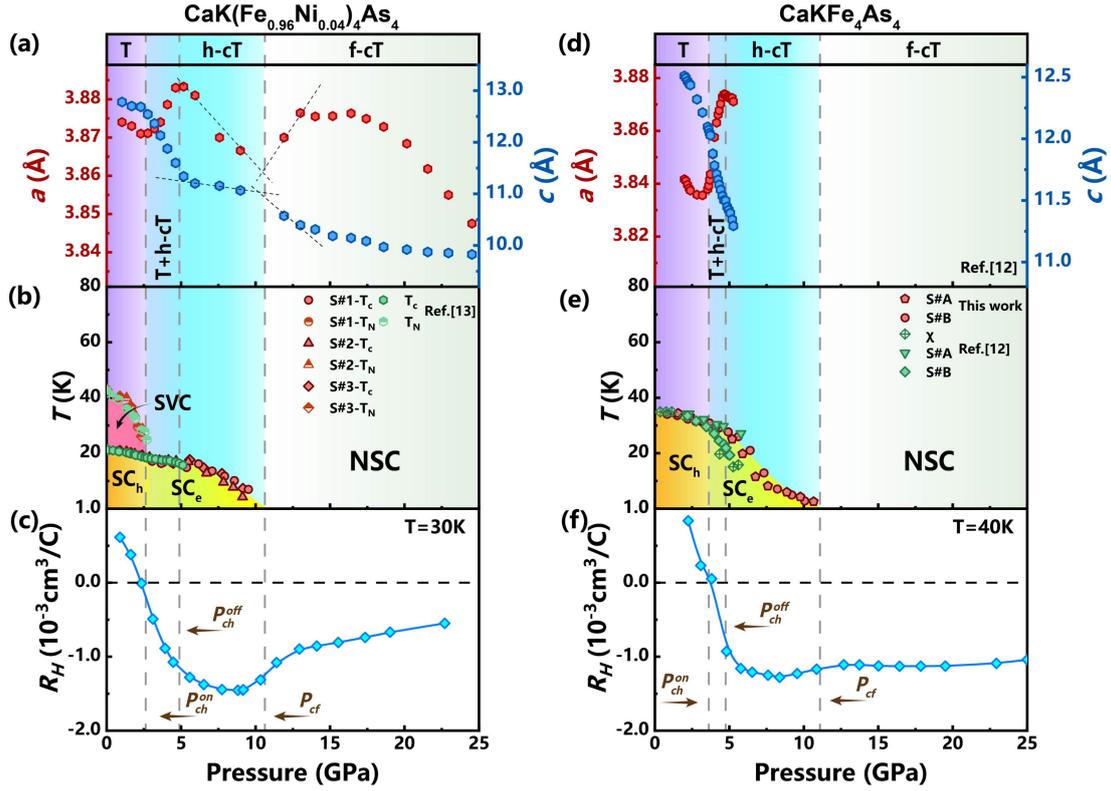

**Figure 4. Structure information, pressure-temperature phase diagram and Hall coefficient ($R_H$) of the CaK(Fe$_{1-x}$Ni$_x$)$_4$As$_4$ (x=0.04 and 0) samples at different pressures.** (a) and (d) Lattice parameters $a$ and $c$ versus pressure for the doped- and undoped- samples. (b) and (e) Pressure-Temperature phase diagrams, displaying the evolution of the spin-vortex crystal (SVC), superconducting (SC) and non-superconducting (NSC) states upon increasing pressure for the doped- and undoped-samples. (c) and (f) Pressure dependence of Hall coefficient ($R_H$) for the CaK(Fe$_{0.96}$Ni$_{0.04}$)$_4$As$_4$ and CaKFe$_4$As$_4$ samples. T, h-cT and f-cT represent the tetragonal phase, half-collapsed tetragonal phase and full-collapsed tetragonal phase, respectively. $T_N$ stands for the onset transition temperature of the SVC state. SC$_h$ and SC$_e$ represent the superconducting state with the dominance of hole-carriers and the superconducting state with the dominance of electron-carriers. The $P_{ch}^{on}$, $P_{ch}^{off}$ represent the turn-on and turn-off pressures for the transition from the T phase to the h-cT phase. The $P_{cf}$ stands for the critical pressure for the transition from the h-cT phase to f-cT phase. The data in the figure (d) are taken from the Ref. [12].